\documentclass[aps,prb,twocolumn,groupedaddress,superscriptaddress,nofootinbib]{revtex4-1}
\usepackage{graphicx}
\usepackage{subfigure}
\usepackage{amsmath}
\pdfoutput=1

\usepackage{array}
\newcolumntype{M}[1]{>{\centering\arraybackslash}m{#1}}

\usepackage{hyperref}
\hypersetup{colorlinks=true, linkcolor=blue, citecolor=blue, urlcolor=blue}

\usepackage[normalem]{ulem}
\usepackage[usenames, dvipsnames]{xcolor}

\begin{document}

\title{Pressure-induced insulator-metal transition in EuMnO$_3$}
\author{R. Qiu}
\affiliation{Physique Th\'eorique des Mat\'eriaux, Universit\'e de Li\`ege (B5a), B-4000 Li\`ege, Belgium}
\affiliation{CNRS, Universit\'e de Bordeaux, ICMCB, UPR 9048, F-33600 Pessac, France}
\author{E. Bousquet}
\affiliation{Physique Th\'eorique des Mat\'eriaux, Universit\'e de Li\`ege (B5a), B-4000 Li\`ege, Belgium}
\author{A. Cano}
\affiliation{CNRS, Universit\'e de Bordeaux, ICMCB, UPR 9048, F-33600 Pessac, France}

\begin{abstract}
We study the influence of external pressure on the electronic and magnetic structure of EuMnO$_3$ from first-principles calculations. We find a pressure-induced insulator-metal transition at which the magnetic order changes from $A$-type antiferromagnetic to ferromagnetic with a strong interplay with Jahn-Teller distortions. In addition, we find that the non-centrosymmetric $E^*$-type antiferromagnetic order can become nearly degenerate with the ferromagnetic ground state in the high-pressure metallic state. This situation can be exploited to promote a magnetically-driven realization of a non-centrosymmetric (ferroelectric-like) metal.  
\end{abstract}
\date{\today}
\maketitle

\section{Introduction}

Manganese-based perovskite oxides are well known for displaying the colossal magnetoresistance (CMR) phenomenon. This intriguing feature is associated to a paramagnetic-insulator to ferromagnetic-metal transition taking place in these systems.
CMR compounds mainly derive from the prototypical perovskite LaMnO$_3$, where the insulator-metal transition can be induced by either doping with divalent cations such as Ca, Sr and Ba \cite{jin94, ramirez97} or external pressure \cite{loa2001pressure,Yamasaki2006pressure, ramos2011bandwidth}.
One the other hand, the rare-earth manganites $R$MnO$_3$ ($R=$ Eu, Gd, Tb, ..., Lu) provide an outstanding subfamily of manganites with a very rich temperature-composition phase diagram \cite{bousquet-rewiew}.
These $R$MnO$_3$ compounds display in particular multiferroicity, a property that holds great promises for a novel generation of spintronic devices and related applications.

In contrast to the CMR manganites, no insulator-metal phase transition has been reported in the multiferroic $R$MnO$_3$ systems so far.
Broadly speaking, the multiferroic $R$MnO$_3$ compounds are found to be insulators whose magnetic ground state can evolve from an $A$-type antiferromagnetic ($A$-AFM) state to spin-spiral order and then to an $E$-type antiferromagnet ($E$-AFM). 
This happens in particular if the effective $R$-ion radius is reduced.
Such a ``chemical-pressure''-induced transformation can be interpreted in terms of enhanced magnetic frustration and its likely competition with biquadratic coupling, which favors non-collinear spiral states and collinear $E$-AFM states respectively \cite{mochizuki2009micro, mochizuki2011theory}.
As a result of this interplay, two prominent realizations of magnetically-induced ferroelectricity can be observed in these systems. 
On one hand, we have the spontaneous electric polarization due to spin spiral order as originally observed in TbMnO$_3$ \cite{kimura-03}.
This is currently understood as due to antisymmetric magnetostriction via the so-called inverse Dzyaloshinskii-Moriya or spin-current mechanism \cite{dagotto-a,nagaosa,mostovoy}.
On the other hand, we also have ferroelectricity linked to collinear $E$-AFM order as observed in HoMnO$_3$ \cite{Munoz2001complex, lorenz}.
In this case, the spontaneous polarization is expected from symmetric magnetostriction terms and is generally much larger than other spin-driven ferroelectrics \cite{dagotto-b}.

Recently, the application of external pressure has been found to have a similar effect to that of ``chemical-pressure" in multiferroic $R$MnO$_3$ \cite{aoyama2014giant, aoyama2015multi}.
The spontaneous polarization of TbMnO$_3$, in particular, has been found to increase dramatically above $\sim 4.6$ GPa, which is interpreted as due to the stabilization of the $E$-AFM order over the initial spiral order of the Mn spins \cite{aoyama2014giant}.
A similar increase of the polarization has subsequently been observed in GdMnO$_3$ and DyMnO$_3$ \cite{aoyama2015multi}.
At the same time, the behavior of the corresponding polarization under magnetic field suggests that the rare-earth magnetic moments can interact with the Mn spins and hence have a substantial interference with their pressure-induced multiferroic properties. 
Motivated by these findings, here we study the effect of pressure on the magnetic order of EuMnO$_3$ from first-principles calculations.

EuMnO$_3$ has the $R$-ion with the largest ionic radius among the multiferroic $R$MnO$_3$ compounds. Interestingly, its magnetic properties clearly emerge from the Mn spins since, unlike the other rare-earth manganite multiferroics, the Eu-ion is in a non-magnetic state. 
Multiferroicity can be induced by e.g. Y doping in this system. 
Thus, as a result of the Y-induced chemical-pressure, the system undergoes the whole sequence of phase transitions $A$-AFM $\leftrightarrow$ spiral state $\leftrightarrow$ $E$-AFM by varying the Y content \cite{mochizuki2009micro, mochizuki2011theory}.
In such view, the application of external pressure can be expected to have a similar effect on this system. 
In this paper we show from first-principles calculations that external pressure has, however, a dramatically different influence on EuMnO$_3$.
Specifically, the application of pressure transforms the $A$-AFM-insulator state directly into a FM-metal. 
This unexpected pressure-induced insulator-to-metal transition, although similar to the observed in CMR LaMnO$_3$, is unprecedented within the multiferroic $R$MnO$_3$ series. 
In addition, we find that the non-centrosymmetric $E^*$-AFM state is also metallic in this system and becomes quasi-degenerate with the FM ground state under pressure.
These features make EuMnO$_3$ an unique compound among the manganites because it behaves differently with respect to physical and ``chemical'' pressure, and hosts a genuinely new type of ferroelectric-like metallic state.
To some extent, EuMnO$_3$ can be regarded as bridging the gap between the CMR and multiferroic manganite compounds.

\section{Methods}

\begin{figure}[t!]
\centering
\includegraphics[width=0.5\textwidth]{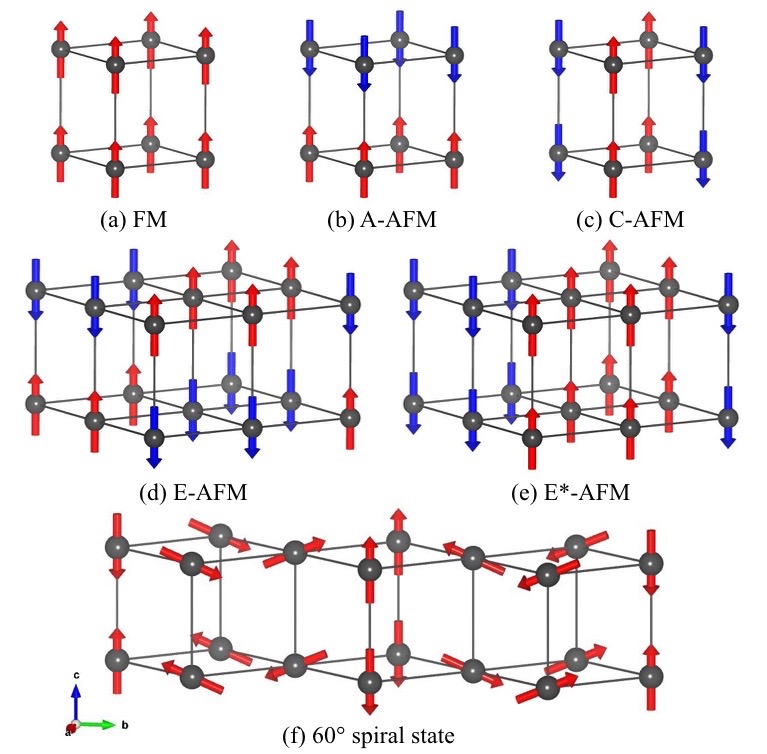}
\caption{Sketch of the different magnetic orders of the Mn spins considered in the text: (a) FM, (b) $A$-AFM, (c) $C$-AFM, (d) $E$-AFM, (e) $E^*$-AFM and (f) 60$^{\circ}$ spiral state (with propagation wavevector along the $b$ axis).} \label{spin_orders}
\end{figure}

Our density functional theory (DFT) based calculations are performed with projected augmented waves (PAW) potentials as implemented in the VASP code \cite{kresse1999paw, kresse1996efficient}. 
We use the generalized gradient approximation (GGA) PBEsol \cite{perdew2008pbesol} exchange correlation functional and apply an on-site Coulomb correction for the Mn-3$d$ states characterized through DFT+$U$ scheme \cite{vladimir1997ldau}.
The Eu-4$f$ electrons are treated as core electrons. 
We consider the most relevant Mn-spin collinear orders found in manganites. 
Namely, ferromagnetic (FM), $A$-, $C$-, $E$- and $E^*$-AFM orders as sketched in figure \ref{spin_orders}.
Note that $E$- and $E^*$-AFM states correspond to the same in-plane Mn spin ordering but with AFM and FM inter-plane coupling respectively [see figure \ref{spin_orders}(d) and \ref{spin_orders}(e)].
In addition, we also consider two representative cases of non-collinear spin-spiral antiferromagnetic order: the 60$^{\circ}$ spiral order with propagation vector $k$ = 1/3 in the $bc$ plane illustrated in figure \ref{spin_orders}(f) and its 90$^{\circ}$ version with $k$ = 1/2 (not shown). 
In our calculations we neglect the spin-orbit coupling. This coupling produces corrections that are at most one order of magnitude smaller than the symmetric exchange interactions (see \textit{e.g.} Ref. \onlinecite{solovyev1996crucial}). Thus, even if it plays a key role for the multiferroic properties (by \textit{e.g.} determining  the value and orientation of the spin-driven electric polarization in the spiral phases \cite{dagotto-a,nagaosa,mostovoy}), it does not introduce qualitative changes in the magnetic phase diagram of the rare-earth manganites \cite{mochizuki2009micro,aoyama2014giant}.
For the collinear orders and the 90$^{\circ}$ spiral order we use an $a \times 2b  \times c$ orthorhombic $Pbnm$ supercell with $6 \times 3 \times 4$ Monkhorst-Pack k-points sampling, 
while the 60$^{\circ}$ spiral configuration is constructed in an $a \times 3b \times c$ supercell using $4 \times 2 \times 3$ k-points grid.
The cutoff energy for plane waves is set at 500 eV. 

\section{Results}

\begin{figure}[b!]
\centering
\includegraphics[width=0.4\textwidth]{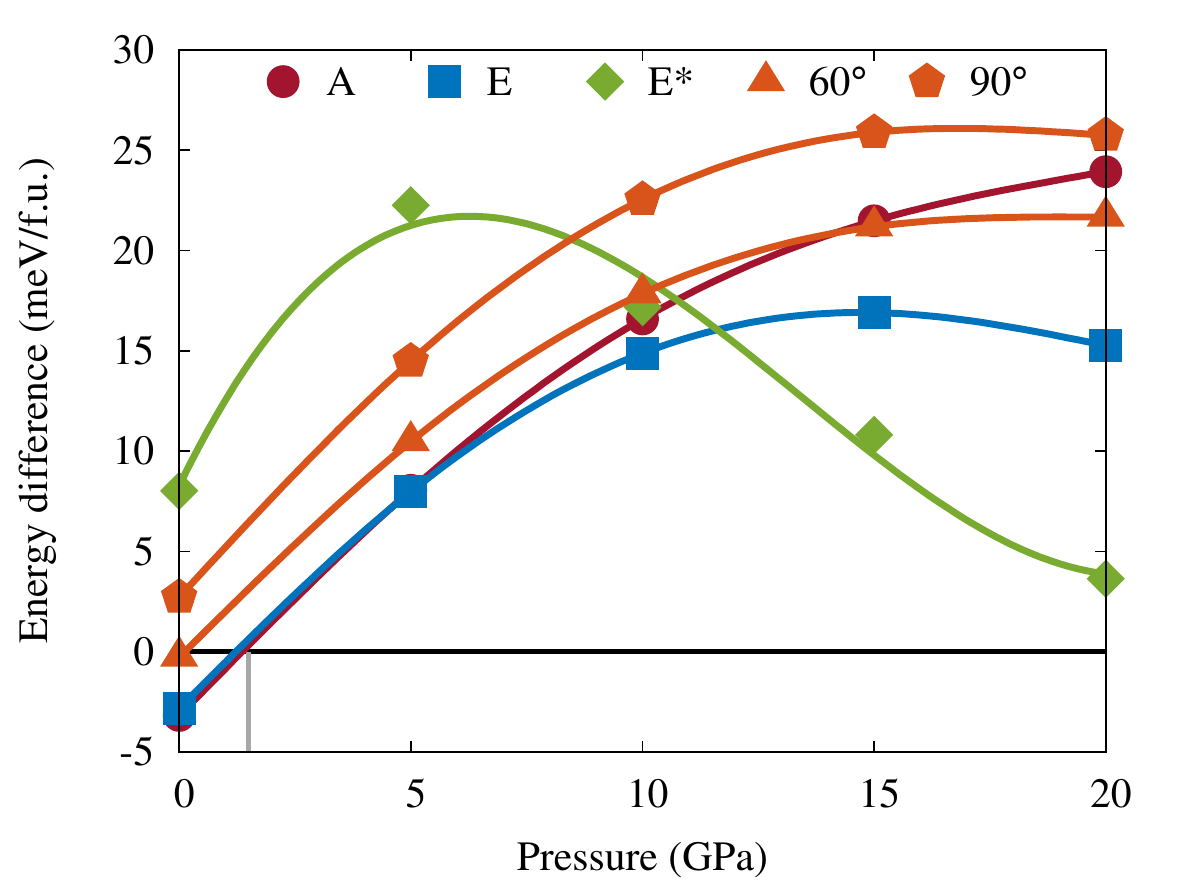}
\caption{Energy of the $A$-AFM, $E$-AFM, $E^*$-AFM 60$^{\circ}$ and 90$^{\circ}$ spiral states as a function of pressure taking the FM state as the reference state. The FM state becomes the ground state at $\sim$ 2 GPa.} \label{u1-e}
\end{figure}

\subsection{A-AFM to FM transition}

In figure \ref{u1-e}, we plot the energy difference between the $A$-AFM, $E$-AFM, $E^*$-AFM, 60$^{\circ}$ and 90$^{\circ}$ spiral states and the FM state as a function of pressure.
The results are obtained by fully relaxing the lattice parameters and internal atomic positions with a Hubbard parameter $U=$ 1 eV. 
We find that the $A$-AFM state has the lowest energy from ambient pressure to $\sim$2 GPa, while the next energy state corresponds to the $E$-AFM order. 
However, by increasing the pressure, the reference FM state eventually has the lowest energy, and hence becomes the ground state of the system. 
We find that the transition between $A$-AFM and FM orders occurs at $\sim$ 2 GPa. 
This transition corresponds to a first-order phase transition in which the net magnetization jumps from 0 to $3.7 \mu_B$/Mn.

Together with this transition, we find that the $E$-AFM order could display a lower energy compared to the $A$-AFM order when the pressure exceeds 5 GPa. 
This is in tune with what is observed in the Tb, Gd and Dy compounds \cite{aoyama2014giant, aoyama2015multi}.
In addition, we observe that, while they can compete with the $E^*$-AFM state at low pressure, both 60$^{\circ}$ and 90$^{\circ}$ spiral orders are always above in energy compared with the FM state.
When it comes to the $E^*$-AFM state, its energy displays an intriguing behavior under pressure. 
As can be seen in figure \ref{u1-e}, the energy of this state shows an important decrease from 5 GPa and tends to the value of the FM state at high pressure ($\Delta E =$ 3.6 meV/f.u. at 20 GPa and further decrease to 2.0 meV/f.u. at 22 GPa).

The zigzag spin-order of the $E^*$-AFM breaks inversion symmetry and transforms the initial $Pbnm$ space-group symmetry of the system into the non-centrosymmetric $Pmn2_1$ one with a spontaneous polar distortion that emerges via symmetric magnetostriction \cite{dagotto-b}. This distortion defines two domains and in principle can be switched by means of its direct link to the $E^*$-AFM underlying structure.
The stabilization of this state then could bring multifuntional properties in EuMnO$_3$ in analogy with the one observed in TbMnO$_3$.
However, according to our calculations, in EuMnO$_3$ the $E^*$-AFM state stays nearly degenerate with the FM state above 20 GPa but it never becomes the ground state of the system.

\subsection{Metallic character of the FM state}

\begin{figure}[t!]
\centering
\includegraphics[width=0.45\textwidth]{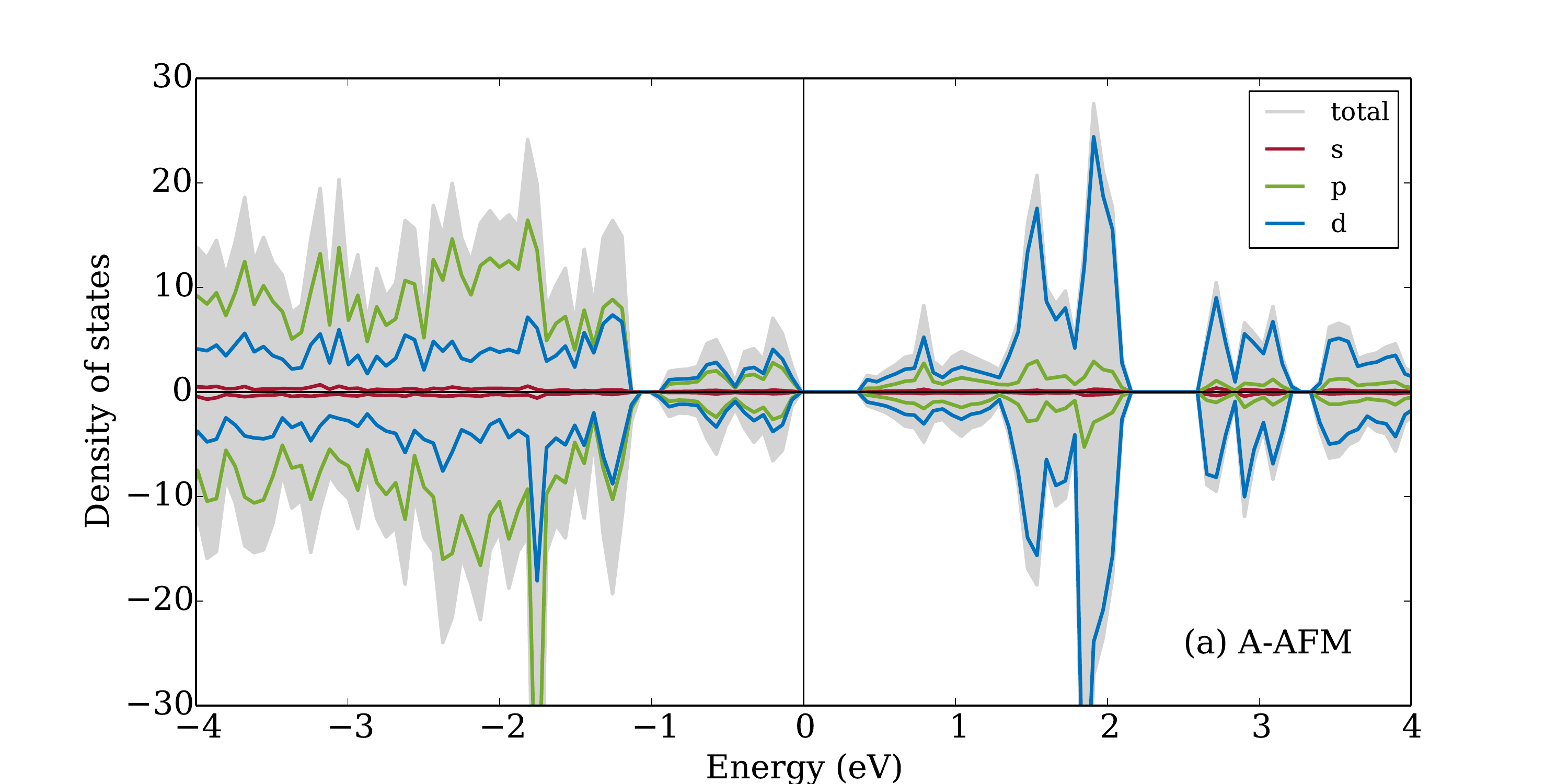}
\includegraphics[width=0.45\textwidth]{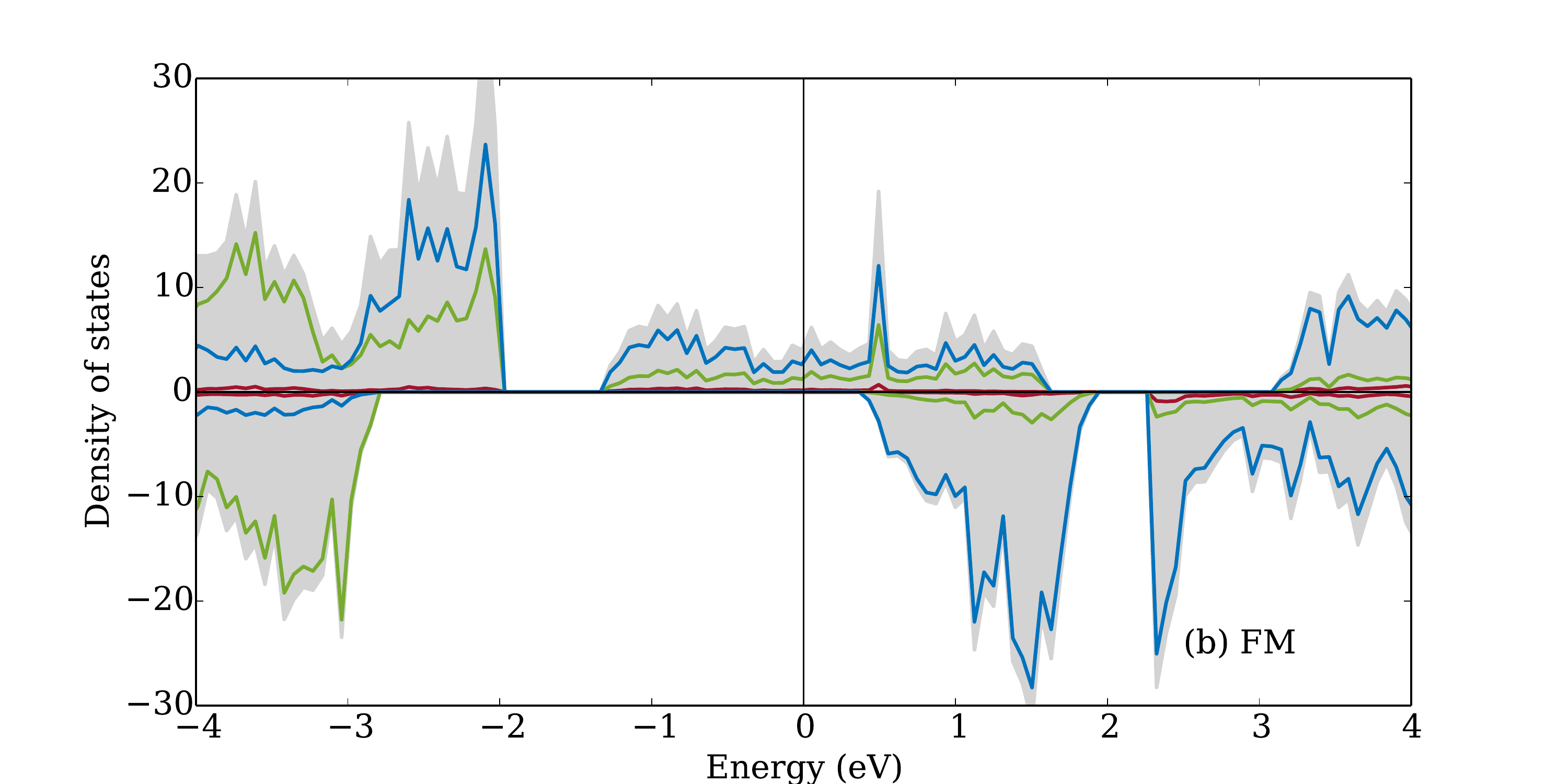}
\includegraphics[width=0.45\textwidth]{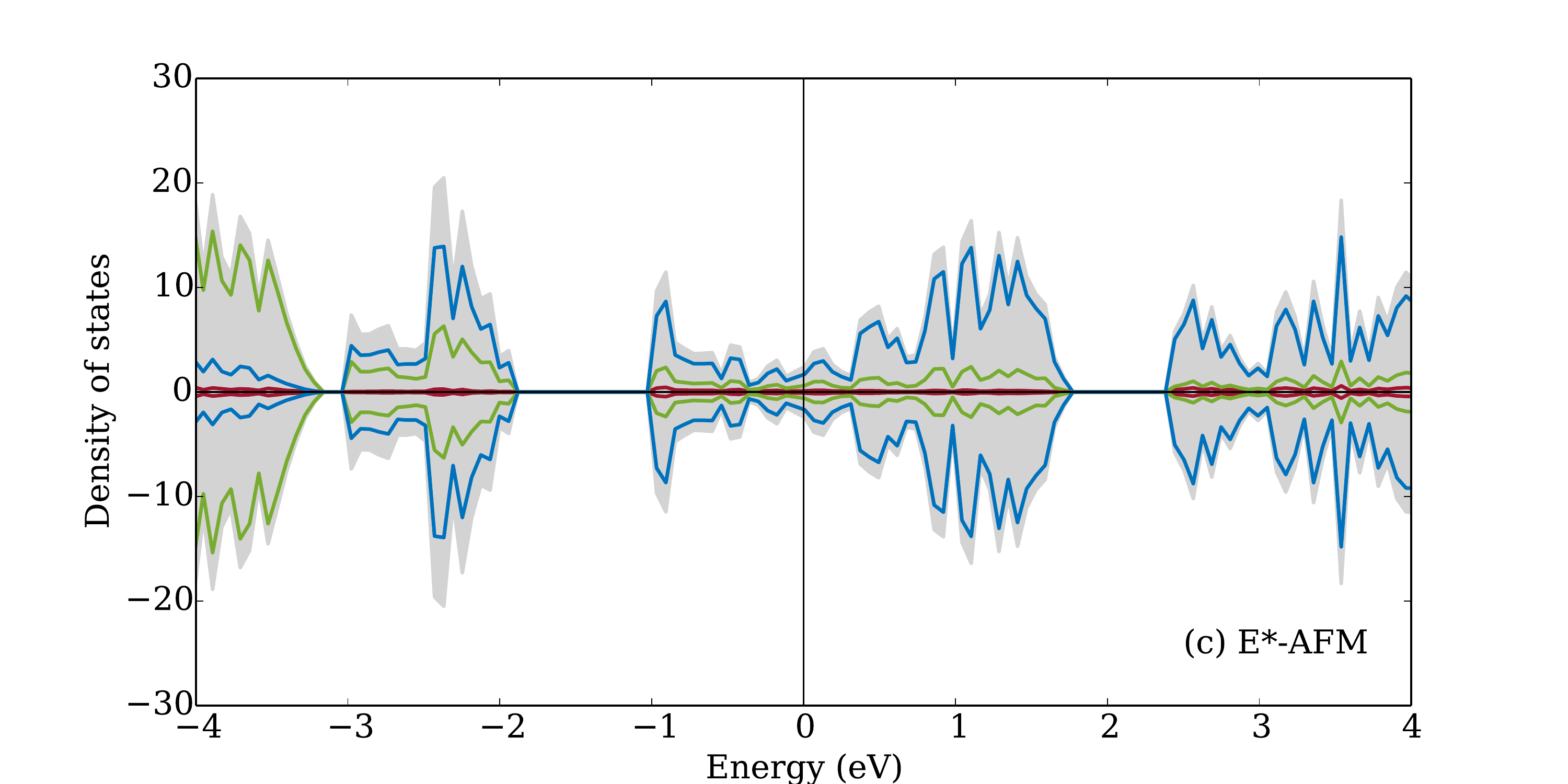}
\caption{Spin-polarized DOS of (a) $A$-AFM (0 GPa), (b) FM (5 GPa) and (c) $E^*$-AFM (20 GPa) states of EuMnO$_3$, where the Fermi level has been shifted to 0 (vertical black line). Total (grey area) and partial ($s$, $p$ and $d$-electrons) DOS are shown, spin-up and -down electrons are mapped on positive and negative area separately. The initial $A$-AFM ground state transforms into the metallic FM state under pressure. The metastable $E^*$-AFM state is also metallic and tends to be nearly degenerate with the FM state at high pressure.}\label{dos}
\end{figure}

In figures \ref{dos}(a) and \ref{dos}(b), we show the density of states (DOS) of the $A$-AFM state at 0 GPa and the FM state at 5 GPa respectively.   
The $A$-AFM DOS displays a gap of 0.5 eV and is symmetric between spin-up and spin-down states. 
The DOS of FM state, on the contrary, has no gap at the Fermi energy for spin-up state, whereas it is gaped for spin-down state.
This finite DOS is dominated by the contribution of Mn-3$d$ orbitals, with a non-negligible contribution of O-2$p$ ones. 
We note that this band structure does not come from a mere shift of the $A$-AFM one, but results from important reconstruction in which structural distortions play a role as we show below. 
Using different values of the $U$ parameter we obtain essentially the same results, and hence we conclude that the FM state in EuMnO$_3$ is therefore a half-metal. 
Thus, we find that the pressure-induced $A$-AFM to FM transition is, in addition, an insulator-metal transition.

In addition, the DOS associated to the $E^*$-AFM order reveals that this state is also metallic as shown in figure \ref{dos}(c). 
In this case, the contribution of the Mn-3$d$ orbitals in the DOS at the Fermi level is even more dominant compared to the FM state. 
Since type of order is accompanied with a polar distortion of the crystal structure that in principle can be switched, the $E^*$-AFM state in EuMnO$_3$ can be seen as an intriguing realization of a magnetically-induced ferroelectric-like metal.

\begin{figure}[t!]
\centering
\includegraphics[width=0.45\textwidth]{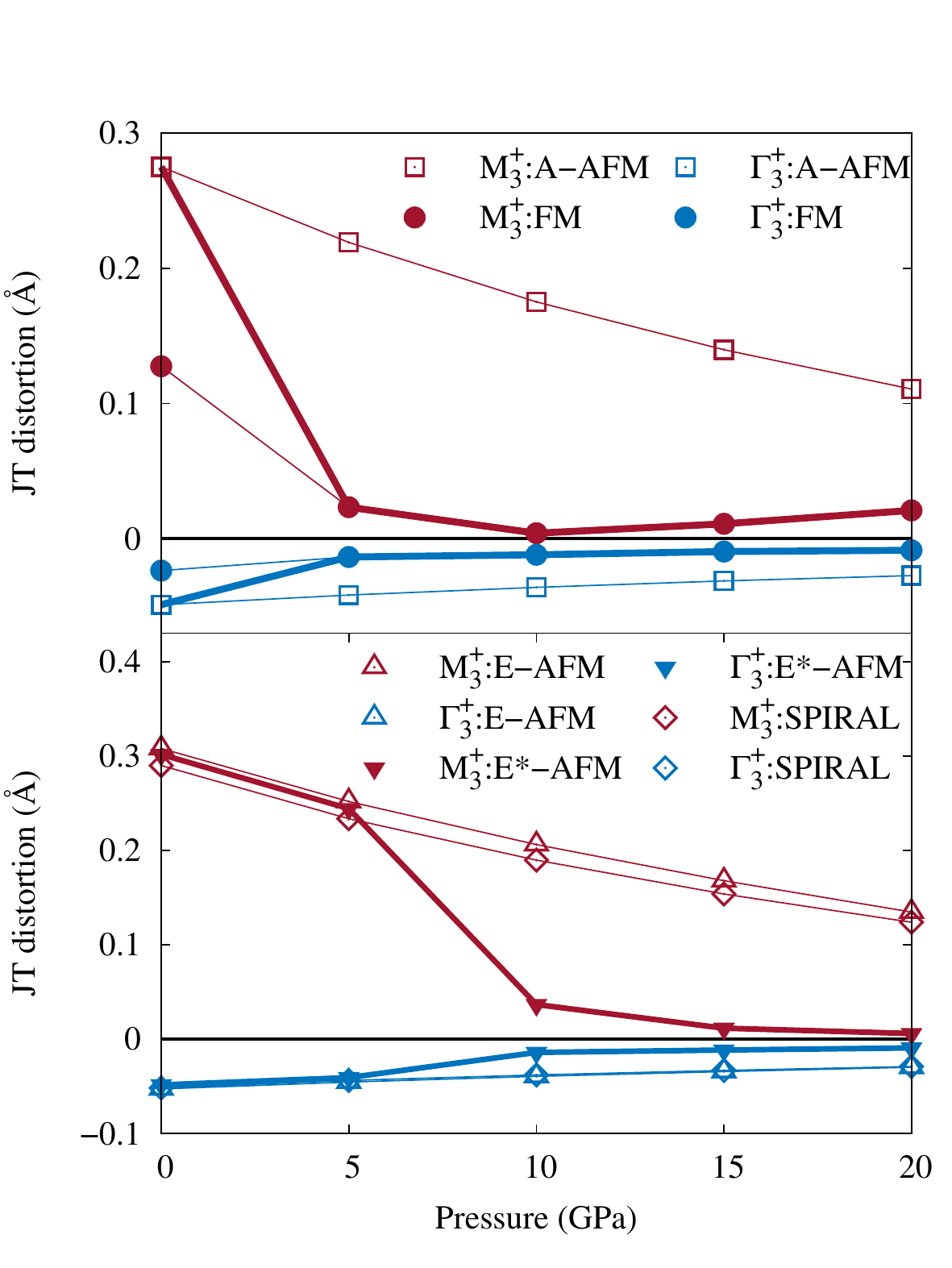}
\caption{Amplitude of the $M_3^+$ (red) and $\Gamma_3^+$ (blue) Jahn-Teller modes as a function of pressure for the different magnetic orders considered in Figure \ref{spin_orders}. Open (close) symbols indicate insulating (metallic) states. The thick lines in the top panel highlight the evolution of the Jahn-Teller distortions in the ground state across the insulator-metal transition. The thick lines in the bottom panel highlight the evolution in the (metastable) $E^*$-AFM metallic state.}\label{dist}
\end{figure} 

\subsection{Interplay between metallicity and Jahn-Teller distortions}

The insulator-metal transition in the reference compound LaMnO$_3$ takes place from a highly Jahn-Teller distorted structure to weakly distorted one and hence is strongly interconnected to the lattice \cite{ramos2011bandwidth,rivero2016a, rivero2016b}.
In order to investigate this aspect in EuMnO$_3$, we performed a symmetry-adapted mode analysis of the distortions that accompany the magnetic orders using the program ISODISTORT \cite{campbell2006isodisplace}.
Thus, we compare the virtual cubic structure with the $Pbnm$ structures obtained for the FM, $A$-AFM and 60$^{\circ}$ spiral orders and the $Pmn2_1$ structures obtained for the $E$-AFM and $E^*$-AFM ones.
All these structures contain Jahn-Teller distortions associated to the $M_3^+$ and the $\Gamma_3^+$ modes  ($Q_2$ and $Q_3$ respectively in the traditional notation, see \textit{e.g.} Ref. \onlinecite{carpenter2009symmetry}). The evolution of these distortions as a function of pressure is shown in Fig. \ref{dist}. 

As we can see, the system displays an abrupt decrease of the Jahn-Teller distortions at the metal-insulator transition due to the different weight of these modes in the $A$-AFM and FM states. Besides, the amplitude of these distortions taken separately decreases for each state by increasing the pressure, which can be interpreted as an increase of the corresponding stiffness. This reduction, however, has a step-like feature for the metallic FM and $E^*$-AFM states while it is gradual for the insulating states.  
This interplay between Janh-Teller distortion and metallicity has indeed a correspondence to the one observed in LaMnO$_3$ (see \textit{e.g.} Refs. \onlinecite{loa2001pressure,Yamasaki2006pressure,ramos2011bandwidth,rivero2016a,rivero2016b}), and hence establishes a parallelism between these two compounds unnoticed so far.

\section{Discussion}

\subsection{Robustness of the first-principles calculations}

Our first-principles calculations suggest that an insulator-to-metal transition can be induced in EuMnO$_3$ by applying external pressure. 
In order to assess the reliability of this prediction, we have carefully analyzed the main premises of these calculations.

First of all, we checked the dependence of the results on the Hubbard $U$ parameter (see Appendix \ref{u-dep}). 
It has been shown that the $U$ correction applied on Mn $d$ orbitals can be taken as zero in other compounds of the $R$MnO$_3$ series such as TbMnO$_3$ \cite{aoyama2014giant}.
In EuMnO$_3$, however, $U=0$ eV gives the $E$-AFM state as the ground state of the system at ambient pressure, and hence is inconsistent with the $A$-AFM state observed experimentally (see table \ref{u0} in Appendix \ref{u-dep}). 
The experimental ground state at ambient pressure is correctly reproduced with $U \ge $ 1 eV. 
Thus, the need of a small but non-zero $U$ parameter in EuMnO$_3$ makes this system a genuinely correlated system compared to other multiferroic manganites. 
Nonetheless, in order to avoid artifacts due to unphysical correlations, we take the lowest possible value of the $U$ parameter that is compatible with the experiments (i.e. $U=$ 1 eV, see Appendix \ref{u-dep}). 

The optimization of the crystal structure turns out to be a crucial point in our calculations. 
To verify our method, we first carried out a comparative study of TbMnO$_3$ and EuMnO$_3$ (see Appendix \ref{tmo-emo}). 
While we reproduce the results reported in Ref. \onlinecite{aoyama2014giant} for TbMnO$_3$, where the authors did their calculation at fixed cell parameters by imposing $A$-AFM order, we however find that these results are strongly affected by structural relaxations. 
The results for EuMnO$_3$, in contrast, are totally robust with respect to structure changes, which supports the predictive power of our calculations. Specifically, the observed competition between spiral and $E$-AFM order in TbMnO$_3$ is captured only by means of the very specific optimization procedure followed in Ref. \onlinecite{aoyama2014giant}, while usual optimization schemes fail. 
This seems to be related to an overestimation of the corresponding magnetostriction couplings and possibly to the interplay between the Mn spins and the additional order of the Tb ones. 
In this respect, EuMnO$_3$ turns out to be a more robust system where the insulator-to-metal transition is always obtained, together with the accompanying changes in the magnetic properties. 
 
\begin{figure}[t!]
\centering
\includegraphics[width=0.42\textwidth]{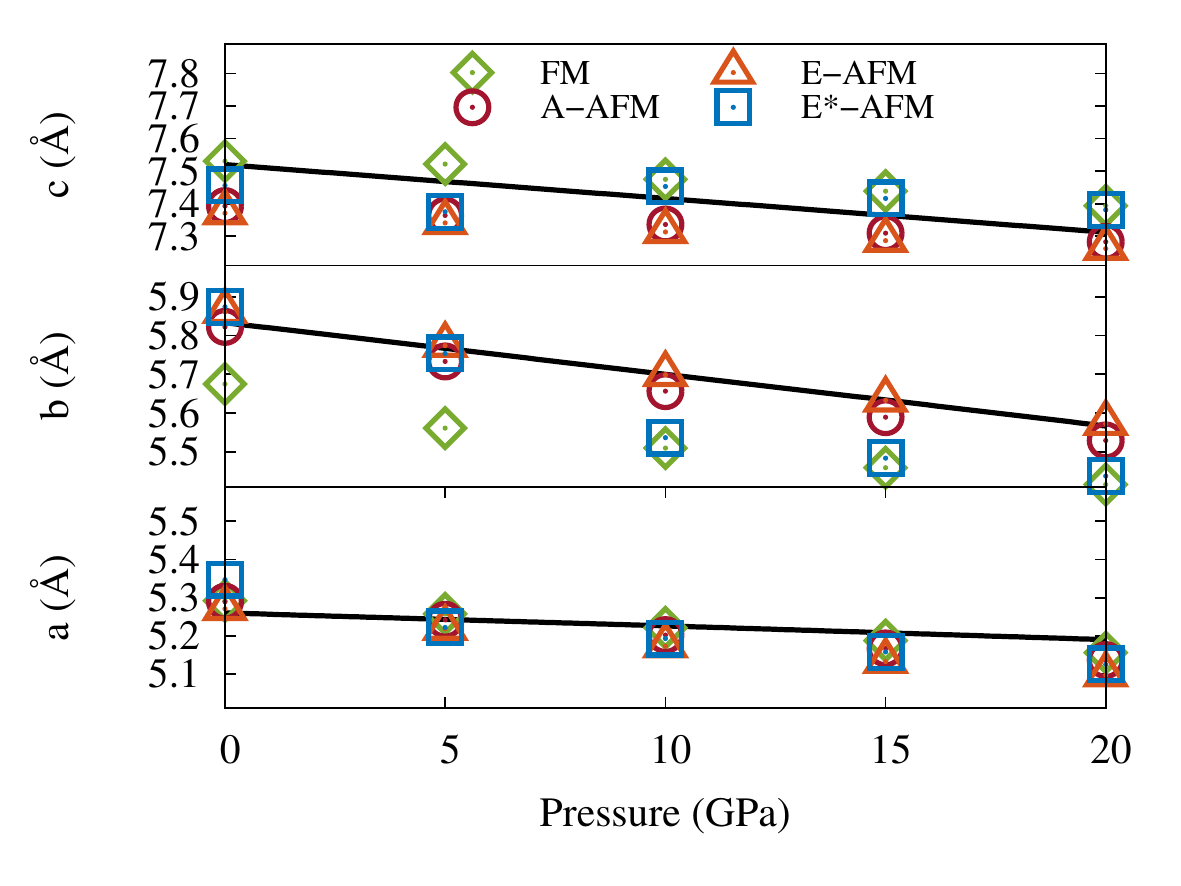}
\caption{Experimental lattice parameters as a function of pressure obtained from Ref. \onlinecite{mota2014dynamic} 
(black lines) and calculated ones for FM, $A$-AFM,  $E$-AFM and $E^*$-AFM orders.\label{lp}}
\end{figure}

The evolution of EuMnO$_3$ under pressure presented in this work has been studied with full atomic and cell relaxations.
The lattice parameters obtained in this way are compared to the experimental data \cite{mota2014dynamic} in figure \ref{lp}.
As we can see, the PBEsol functional produces a good agreement (within a 2\% error) with the experimental data for all the magnetic structures. 
We note that the distortions along $b$ axis are slightly larger in the FM and $E^*$-AFM states, which turns out to be an important parameter to minimize the overall energy. 
Thus, we expect a correct description of the predicted transition at the qualitative level, although the precise value of the e.g. transition pressure has to be taken with a grain of salt.
This is illustrated in our analysis of the dependence of the transition against the $U$ parameter and the structure optimization procedure (see Appendix \ref{u-dep} and \ref{tmo-emo}). 
From this analysis we see that different $U$'s produce different values of the transition pressure, and a similar shift is obtained as a function of the optimization procedure. 
The important point is, however, that the application of external pressure, no matter which calculation procedure we follow, systematically results into a insulator-metal transition in EuMnO$_3$ that, fundamentally, is always the same. This calls for experimental studies of EuMnO$_3$ under pressure to know the exact critical pressure to see the transition.

\subsection{Mapping to a Heisenberg model}

In order to gain additional insight about the microscopic cause of the predicted $A$-AFM-insulator to FM-metal transition, we follow Refs. \onlinecite{aoyama2014giant, fedorova2015biquadratic} map the magnetic energy of the system into a simple Heisenberg model plus a biquadratic coupling term: 
\begin{align}
H &= J_{ab}\sum\limits_{ \langle n,m \rangle }^{ab} {\mathbf S}_n \cdot {\mathbf S}_m 
+ J_{c}\sum\limits_{ \langle n,m \rangle }^{c} {\mathbf S}_n \cdot {\mathbf S}_m 
\nonumber \\
&\quad+ J_{a}\sum\limits_{ \langle\langle n,m \rangle\rangle }^{ab} {\mathbf S}_n \cdot {\mathbf S}_m 
+ J_{b}\sum\limits_{ \langle\langle n,m \rangle\rangle }^{ab} {\mathbf S}_n \cdot {\mathbf S}_m 
\nonumber \\
&\quad
+B\sum\limits_{ \langle n,m \rangle }^{ab} ({\mathbf S}_n \cdot {\mathbf S}_m)^2.
\label{H1}
\end{align}
Here $J_{ab}$ and $J_c$ represent nearest-neighbor interactions in the $ab$ plane and along the $c$ axis respectively, while $J_a$ and $J_b$ are second-nearest-neighbor interactions along $a$ and $b$ respectively.
The biquadratic coupling is restricted to nearest neighbors in the $ab$-plane only and its strength is determined by the $B$ parameter.
The competition between FM nearest- and AFM second-nearest-neighbor interactions is a source of magnetic frustration in the rare-earth manganites. 
This can be quantified by means of the ratio $J_{a(b)}/|J_{ab}|$. 
Thus, the frustration criterion of spiral configuration is 1/2: $J_{a(b)}/|J_{ab}| < 1/2$ favors FM order while $J_{a(b)}/|J_{ab}| > 1/2$ favors the spiral state. 
$J_c$ simply determines if the stacking along $c$ is FM or AFM, while $B \not= 0$ favors collinear orders.

\begin{figure}[t!]
\centering
\includegraphics[width=0.42\textwidth]{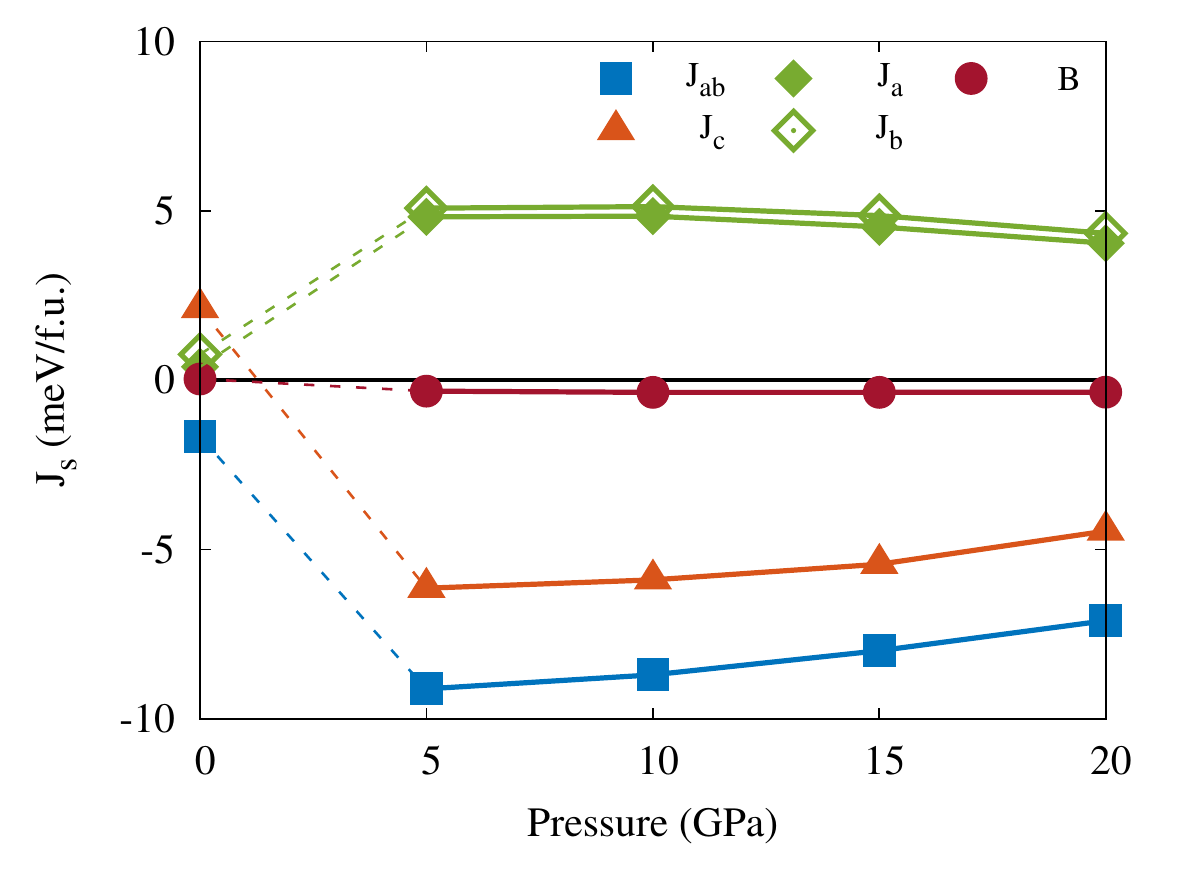}
\caption{Exchange parametres $J_{ab}$, $J_c$, $J_{a}$ and $J_b$ and biquadratic coupling $B$ of the Heisenberg model of Eq. \ref{H1} as a function of pressure. The abrupt change of these parameters at the A-AFM to FM transition is indicated by the dashed line.} \label{phase_diagram}
\end{figure}

In order to determine the parameters of Eq. \ref{H1} in the $Pbnm$ structure, we compute the energy associated to the FM, $A$-, $C$-, 90$^{\circ}$ spiral, and the $E$-AFM sate with the induced polarization along two perpendicular directions ($E_a$- and $E_b$-AFM with $2a \times b \times c$ and $a \times 2b \times c$ supercells respectively) for different pressures between 0 and 20 GPa. In terms of the parameters of the Hamiltonian Eq. \ref{H1}, these energies are
\begin{align}
\begin{split}
E_\text{FM} &= {E_0} + 4(2{J_{ab}} + {J_c}+ {J_a}+ {J_b} + 2B{S^2})S^2,\\
E_\text{$A$-AFM} &= {E_0} + 4(2{J_{ab}} - {J_c}+ {J_a}+ {J_b} + 2B{S^2})S^2
,\\
E_\text{$C$-AFM} &= {E_0} 
+4(-2{J_{ab}} + {J_c}+ {J_a}+ {J_b} + 2B{S^2})S^2
,\\
E_\text{$E_a$-AFM} &= E_0+4(- {J_c} - {J_a}+ {J_b} + 2B{S^2})S^2,\\
E_\text{$E_b$-AFM} & = {E_0} + 4(- {J_c} + {J_a} - {J_b} + 2B{S^2})S^2,\\
E_\text{90$^\circ$spiral} &= {E_0} +4 (- {J_c} + J_a - J_b) {S^2},
\end{split}
\end{align}
where $E_0$ represents the energy of the non-magnetic state. 
In figure \ref{phase_diagram}, we plot the solution of this system of equations as a function of pressure, where the Mn$^{3+}$ spin is taken as $S=2$.

The parameters obtained from this mapping elucidates the intriguing competition between the different magnetic orders in EuMnO$_3$. First of all, we note that the second-nearest-neighbor exchange parameters $J_a$ and $J_b$ are both AFM with a much weaker anisotropy than reported in TbMnO$_3$ \cite{mochizuki2011theory, dong2008origin}. The first-order transition from $A$-AFM to FM state implies the abrupt change of these parameters followed by a more gradual variation. $J_c$, in particular, changes from positive to negative.
In TbMnO$_3$ the biquadratic interaction is enhanced under pressure, which is important for the stabilization of the collinear $E$-AFM phase observed in this system. In EuMnO$_3$, on the contrary, the biquadratic coupling is rather small compared with the exchange interactions at ambient pressure. Furthermore, such a coupling is not enhanced by applying pressure, and therefore is not able to promote the $E$-AFM state. This eventually enables the emergence of the FM order and the accompanying metallicity of the system under pressure.

The mapping to the Heisenberg model, however, has to be taken with some reservations.
If we estimate the N\'eel temperature following a mean-field treatment of the system, we obtain  
$T_N^\text{$A$-AFM} \approx 199$ K (see Appendix \ref{neel}).
The experimental value, however, is 49 K \cite{troyanchuk1997magnetic}. One of the possible reasons of this discrepancy can be related to the metallic character of the FM state itself, as we included this state to compute the $J's$. 
In such a state, the localized-spin picture may not be fully appropriate (even if we find a rather large magnetic moment at the Mn's in the FM state) and/or the exchange interactions can be longer ranged. 
This point requires further investigations that, however, are beyond the scope of the present paper.

\section{Conclusions}
We performed a first-principles investigation of the structural, electronic and magnetic structure of EuMnO$_3$ under pressure.  
We found a pressure-induced insulator-metal transition that is unprecedented in the multiferroic rare-earth manganites $R$MnO$_3$. 
This transition is accompanied with a change of the magnetic order from $A$-AFM to FM, which preempts the spiral and $E$-AFM phases that normally promote multiferroicity in these systems. 
The overall transition, in addition, displays a strong interplay with Jahn-Teller distortions similar to the one observed in LaMnO$_3$. 
EuMnO$_3$ thus establishes an interesting link between colossal-magnetoresistance and multiferroic manganties.
We also found that the non-centrosymetric $E^*$-AFM state is metallic in EuMnO$_3$ and tends to be nearly degenerate with the FM ground state at high pressures.
Thus, EuMnO$_3$ hosts a potential realization of a new type of  (magnetically-induced) ferroelectric metal that can add an extra dimension to the thought-provoking question of ferroelectricity emerging in metals \cite{Anderson1965symmetry, shi2013a, yin2016ferro, kim2016polar, benedek2016ferro}.
These findings are expected to motivate further experimental and theoretical work.

\appendix
\section{\\Dependence on the Hubbard $U$ parameter}\label{u-dep}

In table \ref{u0}, we list the total energy of $A$-AFM and $E$-AFM order by taking FM one as the reference state, calculated with $U=0, 1, 2$ eV at ambient pressure. 
The results show the ground state is $E$-AFM phase for $U=0$ eV, whereas $A$-AFM one for $U = 1, 2$ eV, as we stated in the main text.

\begin{table}
\begin{center}
\begin{tabular}{M{2.5cm}|M{1.2cm}M{1.2cm}M{1.4cm}}
\hline\hline
$U$ value          & FM & $A$-AFM & $E$-AFM \\
\hline
 0 eV         & 0 &      -2.3  & \bf{-18.4}  \\
 1 eV         & 0 & \bf{-3.2} &        -2.8   \\
 2 eV         & 0 & \bf{-4.5} &         4.8    \\
\hline\hline
\end{tabular}
\caption{Total energy of $A$-AFM and $E$-AFM phase with respect to FM one for $U=0, 1, 2$ eV at ambient pressure. }\label{u0}
\end{center}
\end{table}

\begin{figure}[b!]
\centering
\hspace{-0.4\textwidth}(a)\\
\includegraphics[width=0.4\textwidth]{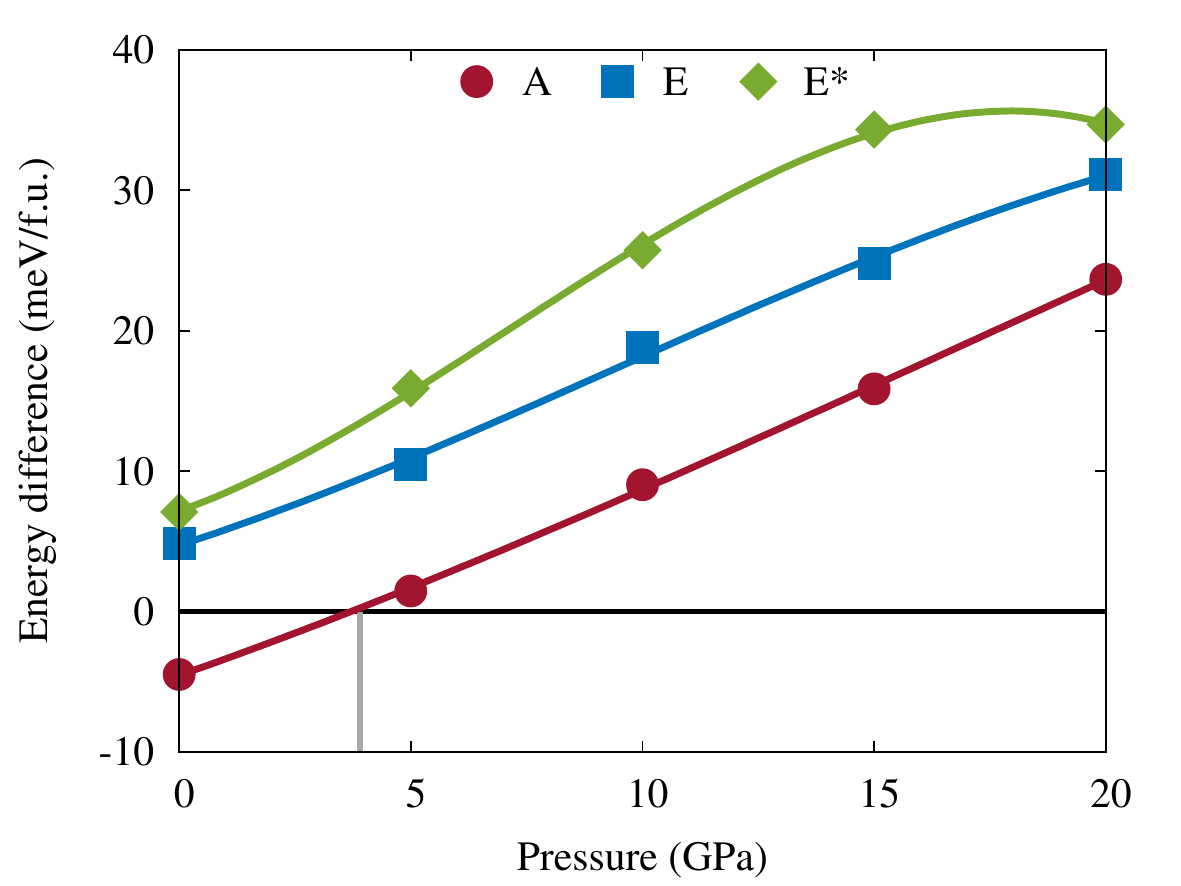}
\\
\hspace{-0.4\textwidth}(b)\\
\includegraphics[width=0.4\textwidth]{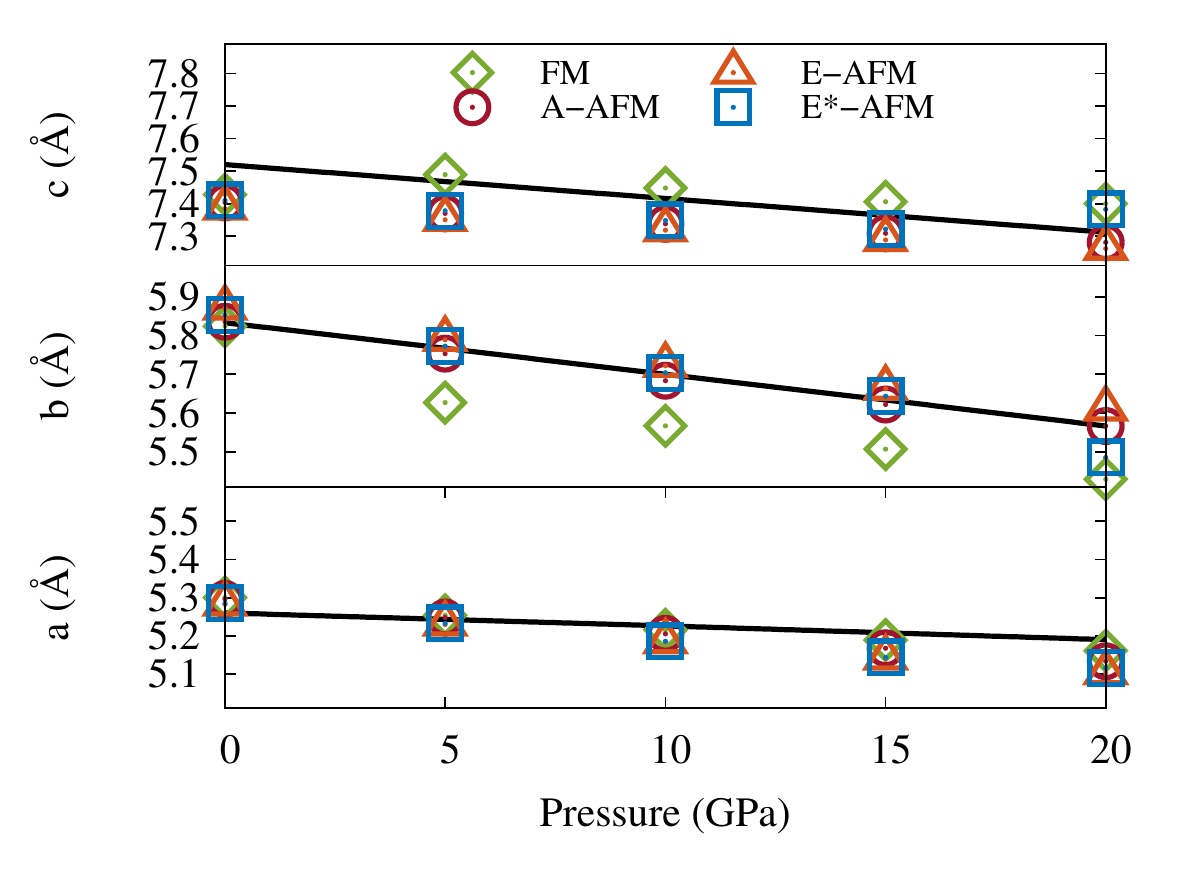}
\caption{(a) Relative energy of the different magnetic orders as a function of pressure for $U=2$ eV. The lattice parameters and the internal atom positions are obtained self-consistently for each magnetic order.
(b) Experimental lattice parameters (black lines) and calculated ones for $U = 2$ eV.}\label{u2}
\end{figure}

In figure \ref{u2}(a) we show the results obtained for $U=2$ eV. 
As for $U=1$ eV, both the lattice parameters and the internal positions are obtained self-consistently for each magnetic state. 
In figure \ref{u2}(a) we see that, compared to the results of $U=1$ eV (figure \ref{u1-e}), the relative energy of the $E$-AFM and $E^*$-AFM states is shifted upwards. 
At the same time, the relative energy between the $A$-AFM order and the FM one remains basically the same and the same crossover is obtained at a slightly higher pressure of $\sim$ 4 GPa. The qualitative picture is thus similar for $U=1$ and $U=2$ eV.
The lattice parameters obtained in this way are compared with the experimental data in figure \ref{u2}(b). 
The degree of agreement is essentially the same as the one obtained for $U=1$ eV (see figure \ref{lp}). 
This confirms that the qualitative prediction of pressure-induced $A$-AFM (insulator) to FM (metal) transition in EuMnO$_3$ is robust with respect to the choice of the $U$ parameter.

\section{\\Dependence on the structure optimization scheme}\label{tmo-emo}

\begin{figure*}[t!]
\centering
TbMnO$_3$ \hspace{200pt} EuMnO$_3$\\
\hspace{-0.4\textwidth}(a)\hspace{40pt}\hspace{0.4\textwidth}(b)\\
\includegraphics[width=0.4\textwidth]{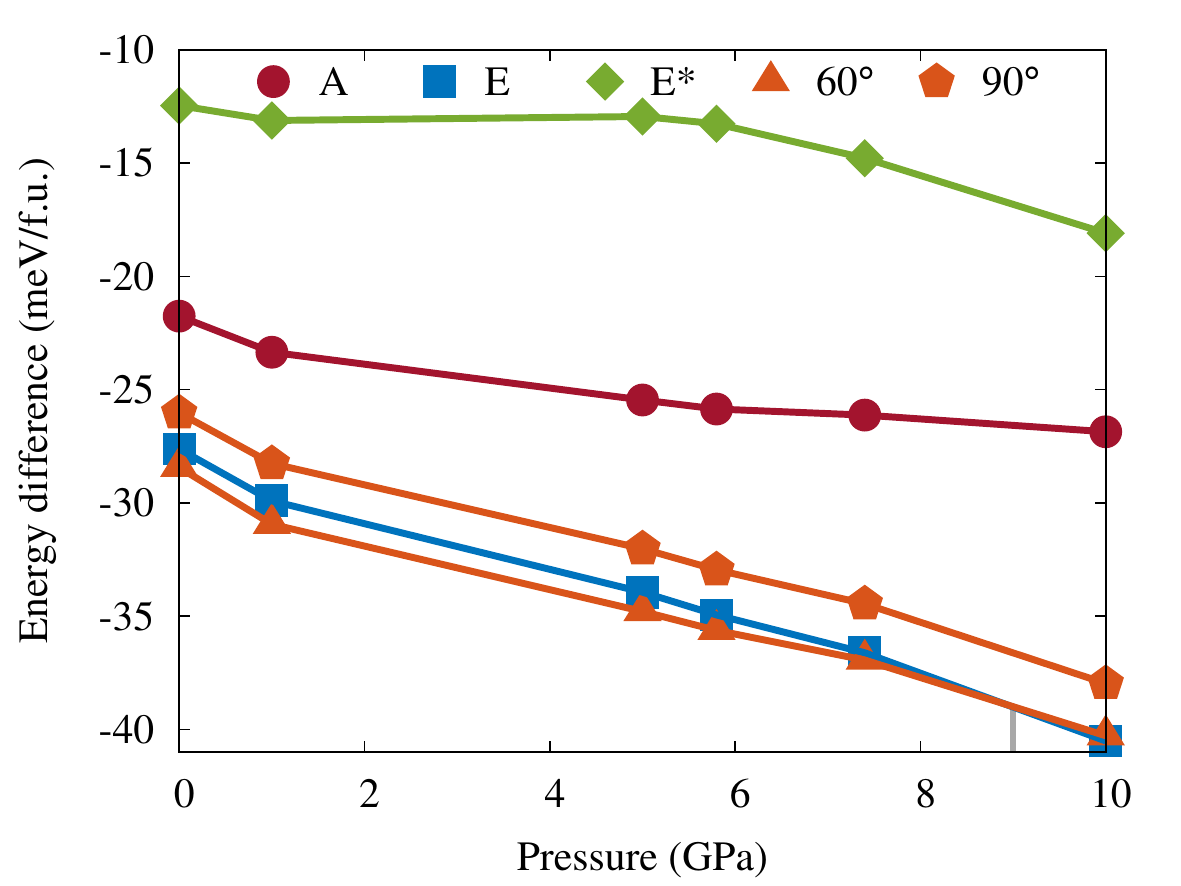}
\hspace{40pt}
\includegraphics[width=0.4\textwidth]{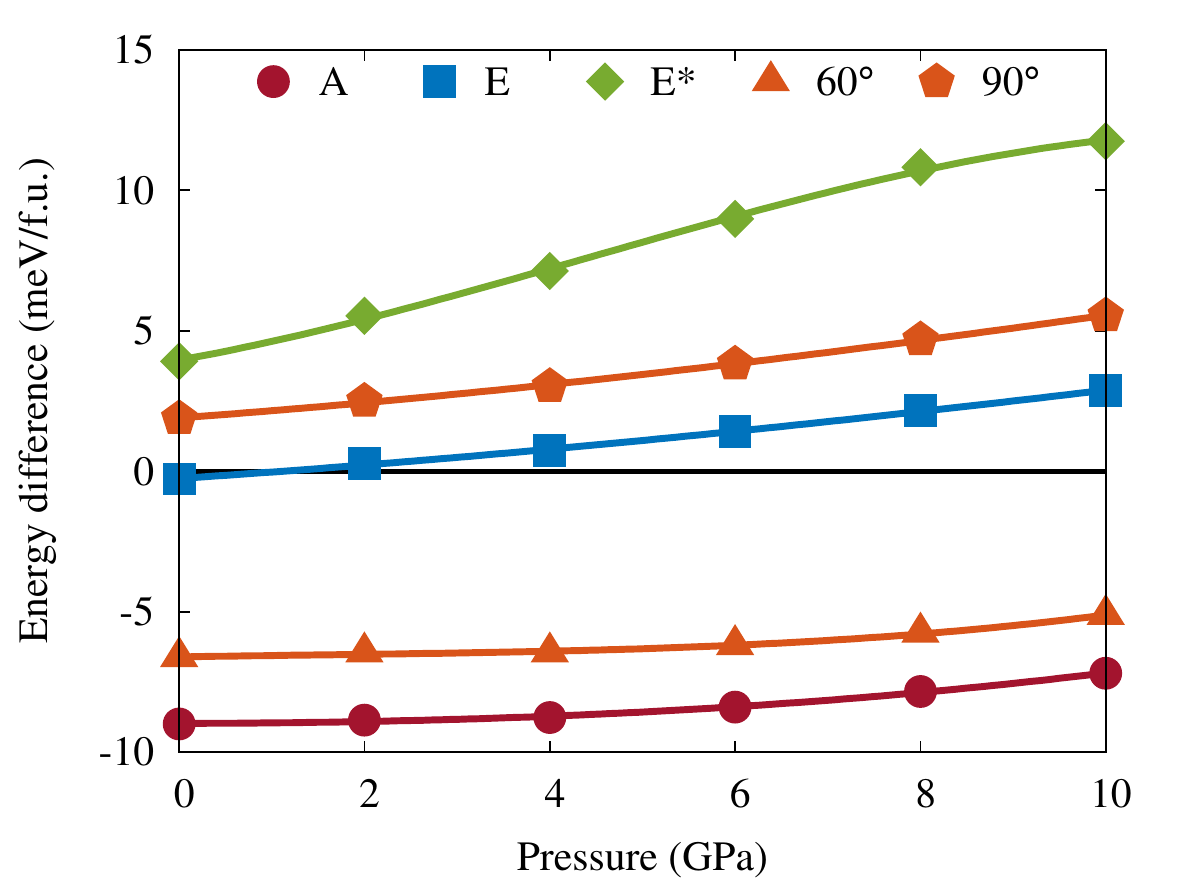}\\
\hspace{-0.4\textwidth}(c)\hspace{40pt}\hspace{0.4\textwidth}(d)\\
\includegraphics[width=0.4\textwidth]{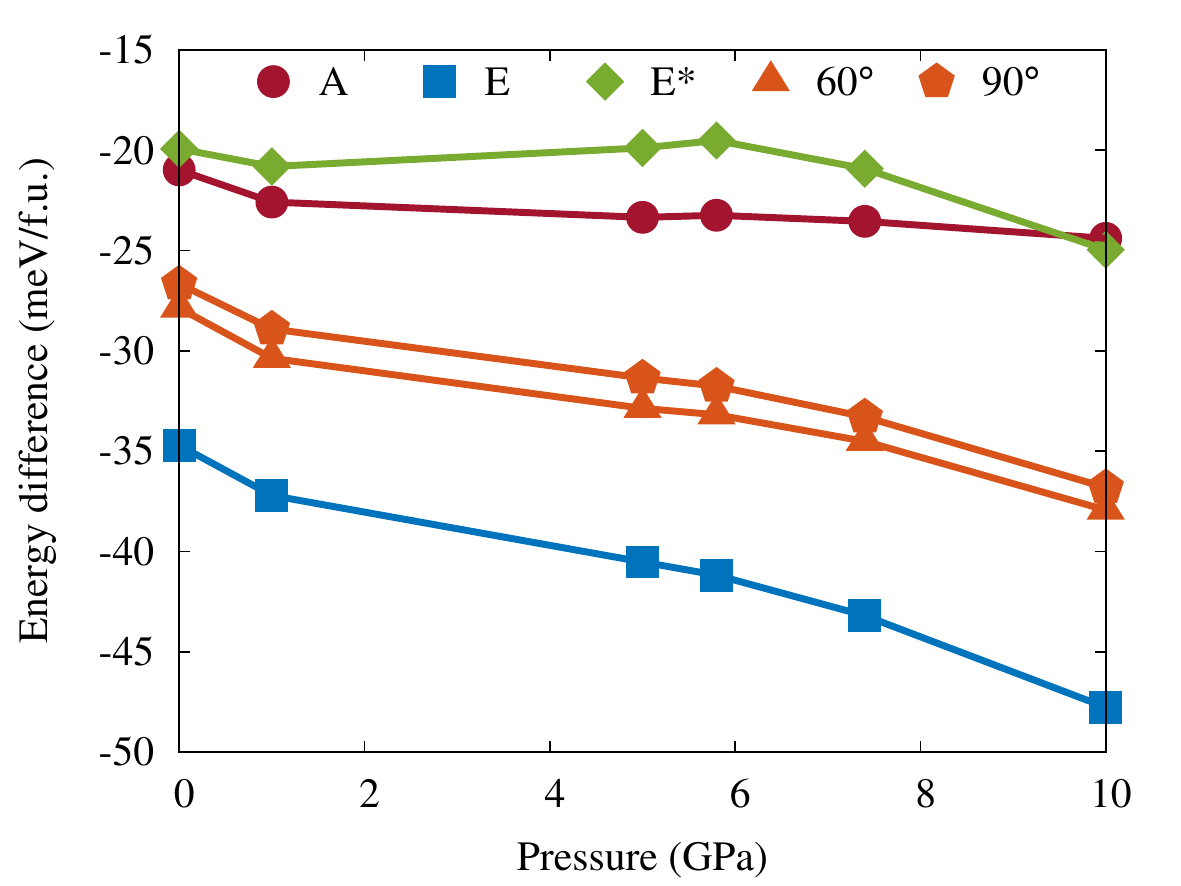}
\hspace{40pt}
\includegraphics[width=0.4\textwidth]{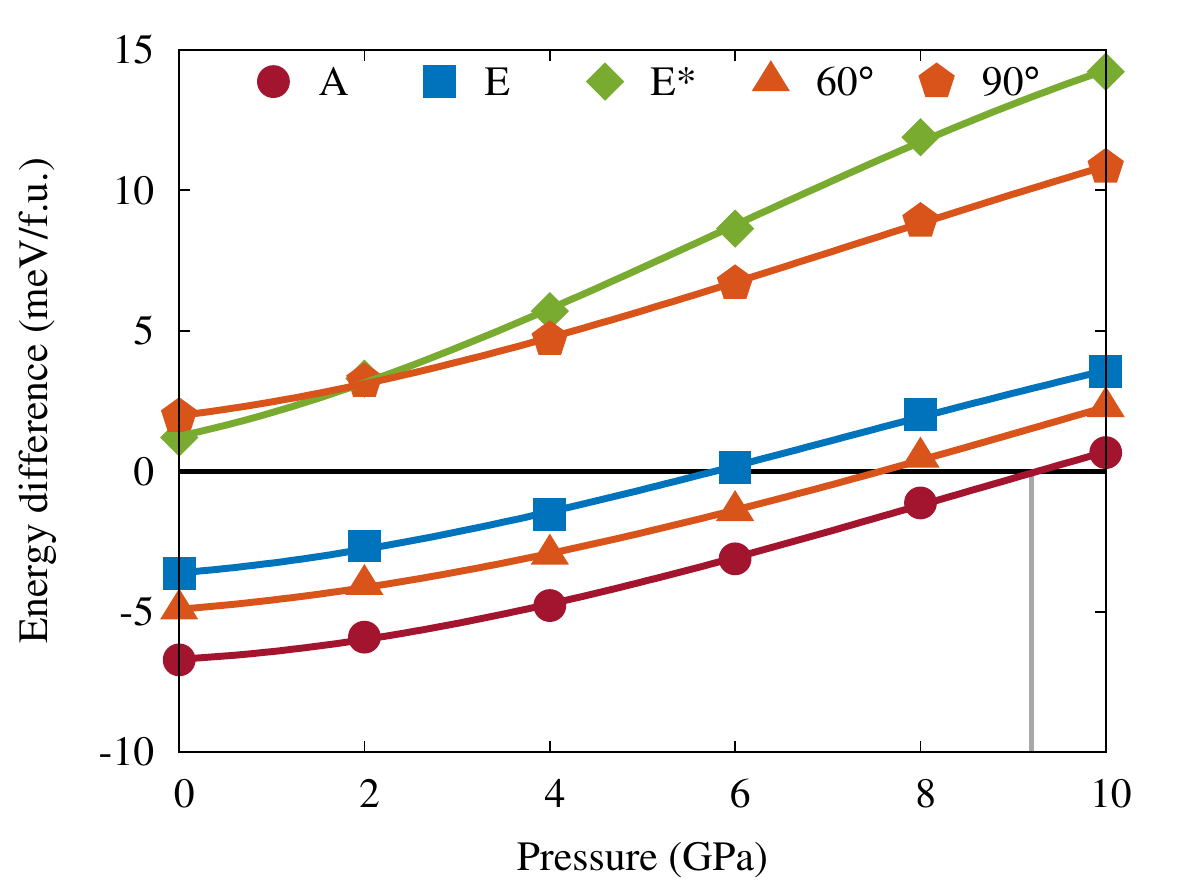}
\caption{Comparative study of the structure optimization procedure in TbMnO$_3$ and EuMnO$_3$. 
In all cases the lattice parameters correspond to their experimental values while the internal positions are obtained following two different methods. 
(a)-(b) $A$-AFM order is imposed and the internal positions are obtained by optimizing the internal coordinates in this magnetic state. 
The output is used to compute the energy associated to the other magnetic orders, with no additional optimization. 
This method is used in Ref. \onlinecite{aoyama2014giant} for TbMnO$_3$, although the $A$-AFM state is  not the ground state of this system. 
(c)-(d) The internal positions are relaxed self-consistently for each type of magnetic order separately. We note the strong sensitivity of the $E$-AFM against the structural relaxations, which changes the qualitative description of TbMnO$_3$ under pressure.
}\label{tmo}
\end{figure*}

In figure \ref{tmo} we compare the results obtained for TbMnO$_3$ and EuMnO$_3$ according to different schemes of structure optimization.  
For TbMnO$_3$ we took $U=0$ eV as in Ref. \onlinecite{aoyama2014giant}. 
For EuMnO$_3$ we took $U=1$ eV to obtain the correct ground state at ambient pressure as explained in the main text. 
In figure \ref{tmo}(a) and \ref{tmo}(b) we plot the results obtained by following the structure optimization described in Ref. \onlinecite{aoyama2014giant}. 
In their paper they relaxed the internal coordinates within the $A$-AFM state at the experimental cell parameters and kept this peculiar relaxed structure fixed to compute and compare the energy of the other magnetic states.
Even if the A-AFM state is never observed to be the ground state in TbMnO$_3$ at any pressure, the results obtained in this way reproduce the experimental transition remarkably well (see figure \ref{tmo}(a) and Ref. \onlinecite{aoyama2014giant}). 
The overestimation of the transition pressure in our calculations could be related to different convergence precision used in Ref. \onlinecite{aoyama2014giant} (2meV/f.u.).
In the case of EuMnO$_3$, if we follow this procedure the $A$-AFM to FM transition occurs at a much higher pressure (not shown in \ref{tmo}(b)). 
Otherwise, as we discussed in the main text, the qualitative picture remains basically the same, besides a shift of critical pressure as a function of the optimization procedure.

In figure \ref{tmo}(c) and \ref{tmo}(d) we show the results obtained according to a more physical procedure of structure optimization. 
In this case the lattice parameters are also fixed to the experimental values, but the internal atomic coordinates are relaxed for each magnetic phase at each value of the pressure. 
This procedure captures magnetostriction effects that are ignored in the previous procedure. 
These effects can indeed be important as they promote \textit{e.g.} the spin-driven spontaneous electric polarization. 
As we see in figure \ref{tmo}(c), this method changes completely the picture in TbMnO$_3$. 
Specifically, among the considered states, the $E$-AFM state becomes the ground state already at zero pressure (while it becomes the ground state beyond 9 GPa if one uses the A-AFM structural parameters). 
Experimentally, however, the ground state corresponds to the spiral order. 
This means that, once magnetostriction effects are switched on, none of the considered spirals reproduce adequately the actual ground state of TbMnO$_3$. 
EuMnO$_3$, in contrast, does not have this complication. 
For this crystal the overall qualitative picture remains the same, even if the energy difference between the different states is now reduced due to the additional energy minimization that comes from magnetostriction effects (see figure \ref{tmo}(d)). 
These magnetostriction couplings then pull the transition pressure down compared to the one obtained according to the procedure of Ref. \onlinecite{aoyama2014giant}. 

For the procedure discussed in the main text, magnetostriction effects are fully taken into account as both lattice parameters and internal positions are relaxed self-consistently for each magnetic state separately. 
This explains the additional shift of the insulator-to-metal transition, and the subsequent possibility of achieving the quasi-degeneracy between FM and $E^*$-AFM states. 

\section{\\Mean-field theory for N\'eel temperature}\label{neel}
We estimate the N\'eel temperature of $A$-AFM using a mean field theory \cite{smart1966effective} based on the exchange parameters $J$'s we obtained from total energy DFT calcuations. Since the anisotropy in the in-plane second-nearest-neighbor interactions is very weak, we simplify this interaction and consider the averaged value $J_2 = (J_a +J_b)/2$ in the following. Thus, we can construct the determinantal equation of the form 
\begin{equation}
\left| {\begin{array}{*{20}{c}}
{\begin{array}{*{20}{c}}
{{a_0}}&{{a_1}}&{{a_2}}&{{a_1}}&{{a_3}}&0&0&0\\
{{a_1}}&{{a_0}}&{{a_1}}&{{a_2}}&0&{{a_3}}&0&0\\
{{a_2}}&{{a_1}}&{{a_0}}&{{a_1}}&0&0&{{a_3}}&0\\
{{a_1}}&{{a_2}}&{{a_1}}&{{a_0}}&0&0&0&{{a_3}}\\
{{a_3}}&0&0&0&{{a_0}}&{{a_1}}&{{a_2}}&{{a_1}}\\
0&{{a_3}}&0&0&{{a_1}}&{{a_0}}&{{a_1}}&{{a_2}}\\
0&0&{{a_3}}&0&{{a_2}}&{{a_1}}&{{a_0}}&{{a_1}}\\
0&0&0&{{a_3}}&{{a_1}}&{{a_2}}&{{a_1}}&{{a_0}}\\
\end{array}}
\end{array}} \right| = 0,
\label{B1}
\end{equation}
for the eight magnetic atoms of the $a \times 2b  \times c$ orthorhombic $Pbnm$ supercell.
Here
\begin{equation}
{a_0} = \frac{{8T}}{C}, {a_1} =  - 4{\gamma _1}, {a_2} =  - 8{\gamma _2}, {a_3} =  - 8{\gamma _3},\label{B2}
\end{equation}
where $T$ is the temperature, $C$ is the Curie constant $C = \frac{NS(S+1)}{3k_B}g^2 \mu_B^2$, and the $\gamma$'s are related to the exchange parameters $J_i$ as 
\begin{equation}
{\gamma_i} = - \frac{{z_{i} J_i}}{Ng^2 \mu_B^2},\label{B4}
\end{equation}
with $J_1 = J_{ab}$ and $J_3 = J_c$.
Among the eight solutions of the Eq. \eqref{B1}, 
\begin{equation}
{a_0} =  - 2{a_1} - {a_2} + {a_3} \label{B5}
\end{equation}
corresponds to the $A$-AFM state. Thus, from Eqs. \eqref{B2} to \eqref{B5} the N\'eel temperature of $A$-AFM state can be estimated as
\begin{equation}
T_N^\text{$A$-AFM} =  {-2S(S+1)\over 3 k_B } \left( 2J_{ab}  + J_a +J_b -  J_c \right). \label{B6}
\end{equation}

\bibliography{References}
\end{document}